\documentclass[aps,reprint,groupedaddress,longbibliography]{revtex4-1}
\usepackage{graphicx}
\usepackage{amssymb}
\usepackage{amsmath}
\usepackage{setspace}

\usepackage[%
  colorlinks=true,
  urlcolor=blue,
  linkcolor=blue,
  citecolor=blue
]{hyperref}
\usepackage{url}

\begin{document}

\title{Multi-axis atom interferometer gyroscope with a single source of atoms}
\date{\today}

\author{Yun-Jhih~Chen\textsuperscript{1,2}}
\author{Azure~Hansen\textsuperscript{1}}
\author{Gregory~W.~Hoth\textsuperscript{1,2}}
\altaffiliation{Present address: Department of Physics, University of Strathclyde, Glasgow, United Kingdom}
\author{Eugene~Ivanov\textsuperscript{1,3}}
\author{Bruno~Pelle\textsuperscript{1,2}}
\altaffiliation{Present address: MUQUANS, Institut d'Optique d'Aquitaine, rue Fran\c{c}ois Mitterrand, 33400, Talence, France. }
\author{John~Kitching\textsuperscript{1}}
\author{Elizabeth~A.~Donley\textsuperscript{1}}

\affiliation{\textsuperscript{1}National Institute of Standards and Technology, Boulder, CO 80305}
\affiliation{\textsuperscript{2}University of Colorado Boulder, Boulder, CO 80309}
\affiliation{\textsuperscript{3}University of Western Australia, Perth, WA, Australia}

\begin{abstract}
Using the technique of point source atom interferometry, we characterize the sensitivity of a multi-axis gyroscope based on free-space Raman interrogation of a single source of cold atoms in a glass vacuum cell. The instrument simultaneously measures the acceleration in the direction of the Raman laser beams and the component of the rotation vector in the plane perpendicular to that direction. We characterize the sensitivities for the magnitude and direction of the rotation vector measurement, which are 0.033~$^{\circ}/\mathrm{s}$ and 0.27~$^{\circ}$ with one second averaging time, respectively. The sensitivity could be improved by increasing the Raman interrogation time, allowing the cold-atom cloud to expand further, correcting the fluctuations in the initial cloud shape, and reducing sources of technical noise. The unique ability of the PSI technique to measure the rotation vector in a plane may permit applications of atom interferometry such as tracking the precession of a rotation vector and gyrocompassing. 

\end{abstract}


\maketitle

\section{Introduction}\label{sec.introduction}

Light pulse atom interferometers may help answer some of the most important questions in fundamental physics because of their extraordinary sensitivity to inertial effects \cite{Peters.1999,Harbe.2005,Wolf.2007,Dimopoulos.2007,Arvanitaki.2008,Bouchendira.2011,Graham.2013,Rosi.2014,Hamilton.2015,Canuel.2018}. They may also have applications in navigation and geodesy because of their long-term stability and accuracy \cite{Durfee.2006,Canuel.2006,Stockton.2011,Dutta.2016,Savoie.2018}. The realization of low size, weight, and power (SWaP) atom interferometers would facilitate their transition from the laboratory to applications in the field \cite{Geiger.2011,Barrett.2016,CAL,MAIUS,Cheiney.2018}. Several groups have demonstrated portable atom interferometer gravimeters for field use \cite{Bidel.2013,Farah.2014,Freier.2016,Bidel.2018,Muquans}.

Light pulse atom interferometer gyroscopes are typically more complex than gravimeters. Multiple cold-atom sources or a four-pulse Raman geometry are often used to distinguish between interferometer phase shifts induced by rotation and by acceleration. The approach with multiple cold-atom sources has been demonstrated with counter-propagating atomic beams \cite{Gustavson.1997,Gustavson.1998,Gustavson.2000,Durfee.2006} and cold-atom clouds \cite{Canuel.2006,T.Muller.2009,Gauguet.2009,Tackmann.2012,Rakholia.2014,Berg.2015,Yao.2018} using the most common beamsplitter-mirror-beamsplitter Raman pulse sequence \cite{Raman}. Since the rotational phase shifts depend on the atom velocity while the acceleration phase shifts do not, the rotation and acceleration measurements are constructed from linear combinations of the signals from the counter-propagating cold atom sources. This approach has been successfully implemented in a compact package with a high data rate \cite{Rakholia.2014}. The approach with a four-pulse ``butterfly" Raman geometry \cite{Canuel.2006,Stockton.2011,Dutta.2016,Wu.2017,Savoie.2018} has been demonstrated using a single atom cloud. In this configuration, the rotational phase shift depends on the magnitude of the acceleration in the direction orthogonal to the rotation axis and the Raman beams, complicating the extraction of pure rotation in a sensor that is not geostationary \cite{Stockton.2011}. Atom interferometers with the above three-pulse or four-pulse configuration measure one axis of rotation. Their multi-axis sensitivity can be achieved by interleaving measurements with Raman beams propagating along different axes and/or with different pulse sequences, as demonstrated in \cite{Canuel.2006,Wu.2017}.  
 
\begin{figure}[!b]
\includegraphics[width=\columnwidth]{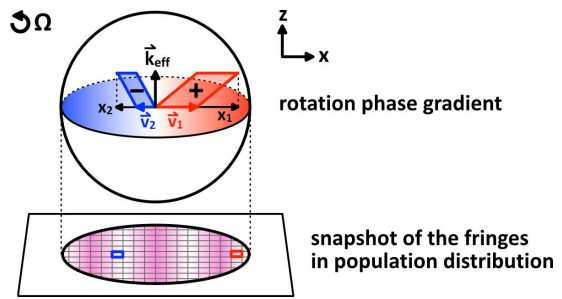}
\caption{(Color online.) Parallel operation of many Mach--Zehnder atom interferometers with a cloud of cold atoms expanding inside a pair of counter-propagating Raman laser beams of relatively large beam diameter. The figure shows two of the many atom interferometers, indicated in red and blue. The position-velocity correlation of a point source preserves the interferometer phase shifts, which are read out as an image of spatial fringe in population distribution.}\label{fig.manyAIs}
\end{figure}
 
Point-source atom interferometry (PSI) is a simultaneous multi-axis gyroscope technique based on a single source of atoms. PSI was previously demonstrated in a 10 m atomic fountain \cite{Dickerson.2013} designed to enable precision tests of the equivalence principle \cite{AIbook.2009}. In PSI, the beamsplitter-mirror-beamsplitter Raman pulse sequence is applied to an isotropically expanding cloud of atoms in which each atom only interferes with itself. Because the enclosed area of the matter-wave Mach--Zehnder interferometer depends on the momentum kick applied to the atom and the atom velocity, the thermal velocity spread of the expanding cloud creates many Mach--Zehnder matter-wave interferometers spanning all directions in a single operation. Each atom acts as an interferometer generating an interferometer phase that depends on the atom's initial velocity as illustrated in Figure~\ref{fig.manyAIs}. The strong position-velocity correlation for atoms in the expanded cloud preserves the phase shifts that are detected as an image. Spatial fringes arising from rotations are imprinted across the cloud on the population distribution between the hyperfine ground states. From the fringe pattern, the acceleration in the propagation direction of the Raman laser beams and the rotation vector components in the plane perpendicular to that direction are measured simultaneously. 

Taking advantage of the dramatic simplifications provided by PSI, we have developed a process amenable to portable applications in which a single cloud of atoms expands and falls by only a few millimeters during an interrogation cycle. We have previously demonstrated rotation measurements with PSI and characterized a systematic error due to the finite size of the  cold-atom cloud \cite{Hoth.2016}. Here we demonstrate the measurement of the acceleration in one direction and the rotation vector in the plane perpendicular to that direction and we evaluate the sensitivity of the rotation vector measurement. 

\section{Simultaneous multi-axis inertial sensing with PSI}\label{sec.background}

\begin{figure}[!b]
\includegraphics[width=\columnwidth]{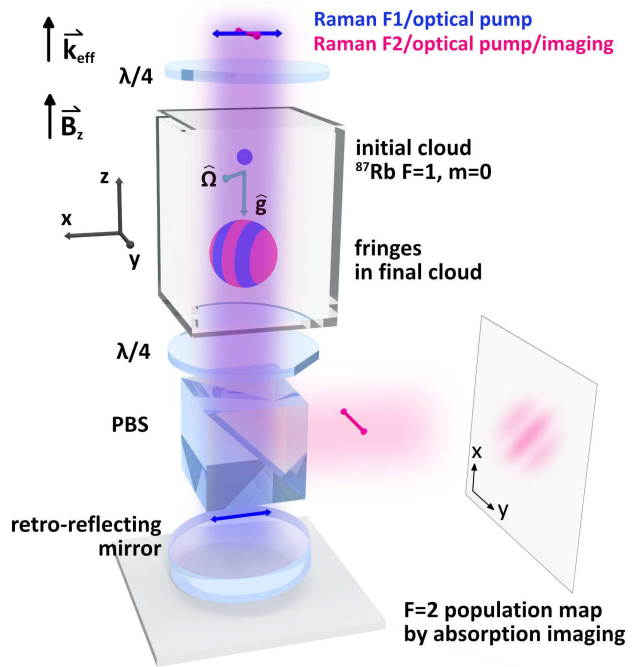}
\caption{(Color online.) Illustration of the glass vacuum cell and surrounding optics of our PSI science package. The inner dimension of the glass cell is 1~cm$^2$. The instrument measures the acceleration in the $z$-direction and the component of the rotation vector in the $xy$-plane.}\label{fig.psi}
\end{figure}

In a ballistically expanding cloud of cold atoms, if the final size of the cloud is much larger than the initial size, referred to as the ``point-source" approximation, the position of an atom is related to its velocity $\mathrm{\mathbf{v}}$ and the expansion time $T_\mathrm{ex}$ of the cloud by $\mathrm{\mathbf{r}}=\mathrm{\mathbf{v}}T_\mathrm{ex}$. In PSI, a beamsplitter-mirror-beamsplitter ($\pi/2-\pi-\pi/2$) pulse sequence is applied to the cloud during its expansion. At the time the cloud is detected, the interferometer phase shift from inertial forces is mapped onto space with the leading terms as
\begin{eqnarray}
\Phi(\mathrm{\mathbf{r}}) & = & \mathrm{\mathbf{k}}_\mathrm{eff} \cdot \mathrm{\mathbf{a}}T^2+
\mathrm{\mathbf{k}}_\Omega\cdot \mathrm{\mathbf{r}}. \label{eqn.phase}
\end{eqnarray}
The first term is the interferometer phase due to a homogeneous acceleration and the second term is a phase gradient across the cloud due to the rotation of the system. The phase gradient is $\mathrm{\mathbf{k}}_\Omega=(\mathrm{\mathbf{k}}_\mathrm{eff}\times\mathbf{\Omega})2T^2/T_\mathrm{ex}$, where ${\mathbf{k}}_\mathrm{eff}$ is the effective Raman transition wavevector, $\mathbf{\Omega}$ is the angular velocity of the system, and $T$ is the time between Raman pulses. 

Since the population ratio is a sinusoidal function of the interferometer phase, the phase gradient across the cloud induced by rotation leads to fringes in the population distribution. The contributions to the interferometer phase from the rotation perpendicular to ${\mathbf{k}}_\mathrm{eff}$ and acceleration in the direction of ${\mathbf{k}}_\mathrm{eff}$ are distinguishable because they modify the fringe pattern in distinct and independent ways. The frequency of the fringes is proportional to the magnitude of the rotation component in the plane perpendicular to ${\mathbf{k}}_\mathrm{eff}$. The orientation of the fringes indicates direction of rotation projected into that plane. The uniform acceleration translates the fringes spatially without affecting the period or the direction of the fringes because the acceleration shifts the overall phase but does not change the phase gradient. 

In our experiment, as illustrated in Figure~\ref{fig.psi}, the Raman laser beams propagate along the $z$-axis and therefore the atoms sense the acceleration in the $z$ direction and the component of rotation in the $xy$-plane. In the point-source approximation, all atoms with the same velocity component in the $xy$-plane have the same final $(x,y)$ position and the same Sagnac area and thus the same interferometer phase shift. We image the cold-atom cloud in the $xy$-plane, which preserves the fringe contrast because signals with the same interferometer phase are integrated along $z$ onto the same pixel in the image. 

When the initial cloud size is not a point source, the rotation fringe contrast and the number of fringes will be lower than that of an ideal point source \cite{Hoth.2016}. A finite-sized cold-atom cloud can be considered as a collection of many point sources. Mathematically, the fringe pattern will be the convolution of this initial cloud shape and the fringe pattern of a point source. As a result of the convolution, fringe contrast is reduced, the number of fringes is fewer, and the direction of the fringes is modified by the initial cloud shape. In a special case where the cloud follows a Gaussian distribution in space, the $x$ and $y$ components of the phase gradient $\mathrm{\mathbf{k}}_\Omega$ are reduced as \cite{Hoth.2016}
\begin{align}
k'_{\Omega,x}&=k_{\Omega,x}(1-\sigma_{0x}^2/\sigma_{fx}^2)\nonumber\\
k'_{\Omega,y}&=k_{\Omega,y}(1-\sigma_{0y}^2/\sigma_{fy}^2)\label{eqn:kp},
\end{align}
where $k_{\Omega,x}=-2k_{\mathrm{eff}}T^2\Omega_y/T_\mathrm{ex}$ and $k_{\Omega,y}=2k_{\mathrm{eff}}T^2\Omega_x/T_\mathrm{ex}$ are the $x$ and $y$ components of the phase gradient for a point source, respectively. The expansion time $T_\mathrm{ex}$ is measured from the time when the cold-atom cloud is released from the MOT or molasses to the time when it is imaged. The terms $\sigma_{0x}$ and $\sigma_{0y}$ are the standard deviation of the initial Gaussian cloud in the $xy$-plane and $\sigma_{fx}$ and $\sigma_{fy}$ are the standard deviations when the cloud is imaged. 
 
\section{PSI with compact science package}\label{sec.psi}

In our PSI science package (Figure~\ref{fig.psi}), the laser beams for state preparation, Raman interrogation, and detection propagate vertically in a shared beam path with an $e^{-2}$ beam diameter of 8~mm and are circularly polarized inside the glass cell. Three orthogonal and retro-reflected beams (not shown in the figure) with $e^{-2}$ diameters of 6~mm form the magneto-optical trap (MOT) for \textsuperscript{87}Rb. The cooling and repumping laser powers are 5.6 and 1.3~mW in each beam, respectively. The laser system that generates the laser beams for the experiment is described in the supplement~\cite{supp}.

Each experimental cycle starts with a \textsuperscript{87}Rb MOT and optical molasses which prepare a cloud of about 10~million atoms. The cloud temperature is about 6~$\mu$K and the standard deviation of the Gaussian initial cloud shape is 0.3~mm. The loading and cooling phases of each cycle are 85~\% of the total cycle when the experimental repetition rate is 5~Hz (and 70~\% at 10 Hz). We apply a 0.7~G quantization magnetic field in the $z$-axis for the following state preparation and Raman interrogation. The atoms are prepared in the $F=1$, $m_F=0$ state in three steps using the lasers from the vertical beam path. First, a 60~$\mu$s pulse of $F=2 \rightarrow F'=2$ transition light pumps the atoms to the $F=1$ state. Second, a 30~$\mu$s pulse of retro-reflected $\sigma^+$ and $\sigma^-$ polarized $F=1\rightarrow F'=0$ light pumps the atoms to the $m_F=0$ sublevel. Third, a 40~$\mu$s pulse of $F=2 \rightarrow F'=3$ light removes the atoms left in $F=2$. The final population in the $F=1$, $m_F=0$ state is at least 85~\%.

After state preparation, we apply Raman pulses with $T=8$~ms and $\pi$-pulse time $t_\pi=4.4~\mu$s. The Raman laser frequencies are tuned 695.815~MHz below the $F'=3$ level in the $5P$ manifold of hyperfine states. The maximum separation of the wave packets is about $94~\mu$m and the enclosed Sagnac area, calculated with the root-mean-square (RMS) velocity of the atoms, is about 0.03~mm$^2$. We delay the Raman laser pulse sequence until 8~ms after the end of the molasses stage so that the desired Doppler-sensitive and magnetically-insensitive Raman transitions are well resolved from unwanted Raman transitions (which are excited due to the finite polarization extinction ratio and/or spurious reflections of the Raman beams).  The power in each Raman laser beam is adjusted to minimize the AC Stark shift of the Raman transition frequency. The cold-atom cloud falls about 3~mm in each experimental cycle.

The rotation fringes are imprinted in both \textsuperscript{87}Rb hyperfine ground states and the patterns are complementary to each other. Detecting either state is sufficient and the population ratio is not required, but doing so would increase the signal-to-noise ratio if the difference in the Gaussian cloud shapes, due to imaging delays or optical pushing, could be balanced. In our case, the initial state of the atoms is $F=1$, $m_F=0$ and for technical reasons only the atoms transferred to the $F=2$ state after the Raman pulse sequence are imaged. We take an absorption image of the atoms in $F=2$ with a 5~$\mu$s pulse of  $F=2\rightarrow F'=3$ probe laser light, pump the atoms to the $F=1$ non-imaged state with 2~ms of the MOT cooling light, and then take an image of the probe laser light field.  From those two images, the density of the final cloud in the $F=2$ state integrated along the $z$-axis is calculated. Typically, there are two million atoms in the $F=2$ state at detection. 

\section{Rotation fringe detection}\label{sec.pca}

\begin{figure}[!b]
\includegraphics[width=3.1in]{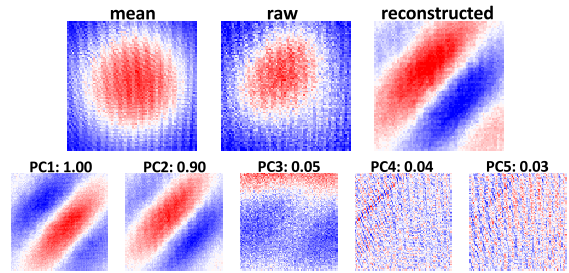}
\caption{(Color online.) PCA example for a set of 201 raw absorption images. Images are autoscaled with red indicating the F=2 population maxima and blue indicating the minima. Top row: mean image, the first image of the data set, the first image reconstructed with only PC~1 and PC~2 (mean image subtracted). Because of the Gaussian cloud shape and other noise in the image, it is difficult both to see the rotation fringes in the raw image and to do a 2D fit to the fringes. Bottom row: the first five PCs and their variances. The variances are normalized to the value of PC~1. The image size is $2.1\times2.1$ mm$^2$ (80$\times$80 pixels). The simulated rotation is 2.94~$^{\circ}$/s in the diagonal direction. The average of the rotation rates measured from the reconstructed images is 2.70~$^{\circ}$/s, which is lower than the applied rate because of the finite size of the initial cloud~\cite{Hoth.2016}. The rotation fringe contrast is about 50~\% after we process the images with PCA.}\label{fig.pca}
\end{figure}

We characterize the PSI gyroscope with simulated rotations by tilting the retro-reflecting mirror in Figure~\ref{fig.psi} during the Raman interrogation, which is formally equivalent to rotating the entire system \cite{Lan.2012}. The mirror is mounted on a piezo-actuated platform, whose $x$- and $y$-angular displacements are controlled via analog voltages. The simulated rotation is about $10^3$ times larger than the Earth rotation and generates about one fringe period across the final cloud.

In our setup, the direction of the Raman laser beams is in the direction of the local gravity vector $\mathbf{g}$ so the frequency difference between the Raman lasers is chirped to compensate for the Doppler shift due to the free fall of the atoms. With this frequency chirp, the acceleration phase is $(k_{\mathrm{eff}}g-2\pi\alpha)T^2$ \cite{Cheinet.2006}, where $\alpha$ is the chirp rate. As a consequence, scanning the chirp rate $\alpha$ provides a convenient way to scan the acceleration phase. 

In this work, we detect the rotation fringe images from the raw absorption images with principal component analysis (PCA) \cite{Shlens.2014,Segal.2010}. PCA has been widely used in machine learning. A few examples of PCA applications in cold-atom physics can be found in references \cite{Dickerson.2013, Segal.2010, Dubessy.2014, Ferrier-Barbut.2018}. In our case,  PCA can be considered an imaging analogy of lock-in detection. We scan the acceleration phase by scanning the chirp rate $\alpha$ so that the rotation fringes translate from shot to shot. The PCA algorithm identifies the moving components such as the rotation fringes from the raw images.

To use PCA in our experiment, we record a series of images at different chirp rates. The images are averaged to create a mean image in which the moving fringes are washed out but the envelope of the cold-atom cloud is retained. The mean image is then subtracted from each image to create a set of zero-mean images. These zero-mean images are the input to the PCA algorithm, which returns two main outputs: (1) a set of basis images called principal components (PCs) and (2) the variance for each PC, which is the variance of the projection of each input image into that PC. Figure~\ref{fig.pca} shows an example of PCA. Because the rotation fringes move as we scan the acceleration phase, the PCA returns a pair of principal components, PC~1 and PC~2, that have the same period and orientation as the rotation fringes but are 90~$^{\circ}$ out of phase (sine-like and cosine-like). The linear combination of PC~1 and PC~2, with their projections into each raw image varying from frame to frame, recreates the moving fringes; as a result, they have the highest variances of the projections. The other PCs, for example, the thin stripes caused by our imaging system (PC 4 and 5), do not follow the scanning of the acceleration phase; they are relatively static in the images and have smaller variances. In general, features of interest can be enhanced by reconstructing the images using only the PCs with the largest variances when we intentionally perturb the system, such as by scanning the acceleration phase.

We reconstruct the fringe images with only the sine-like and cosine-like PCs and disregard all the other PCs of lower variances. The reconstructed and zero-mean images are 2D fitted with $n(x,y)=A\exp(-x^2/2\sigma^2_x-y^2/2\sigma^2_y)\cos(k_xx+k_yy+\phi)$, where $A,~\sigma_x,~\sigma_y,~k_x,~k_y$, and $\phi$ are fit coefficients. We call $\phi$ the offset phase of the rotation fringes. The components of rotation in the $x$ and $y$ directions are calculated as $\Omega_x=k_yT_\mathrm{ex}/(2k_{\mathrm{eff}}T^2)$ and $\Omega_y=-k_xT_\mathrm{ex}/(2k_{\mathrm{eff}}T^2)$. The rotation rate is $\surd(\Omega_x^2+\Omega_y^2)$ and the rotation direction is $\tan^{-1}(\Omega_y/\Omega_x)$. 

\section{Acceleration and Rotation with PSI}\label{sec.cw.ccw}

\begin{figure}[!b]
\includegraphics[width=3.1in]{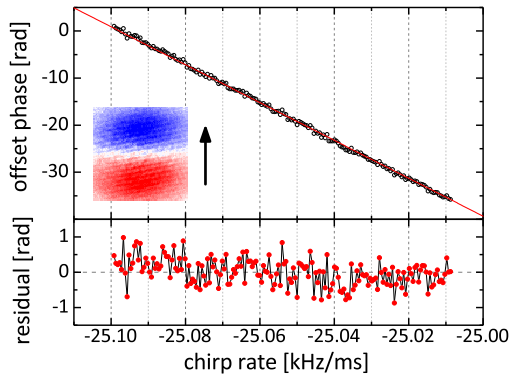}
\caption{(Color online.) Offset phase of the rotation fringes as the acceleration phase is scanned by scanning the Raman laser chirp rate. The inset of a $2.1\times2.1$ mm$^2$ area shows the first fringe image (zero-mean, reconstructed) in the data set of 181 images. The arrow indicates the direction of the fringes' travel. The red line is a linear fit with fixed slope $-402.124$ rad ms/kHz. The simulated rotation is 2.02~$^{\circ}$/s in the $+x$ direction. The average measured rotation rate is 1.60~$^{\circ}$/s. The experimental repetition rate is 5~Hz and the duration of the data set is 36~s. The slow drift in the residual plot is likely due to the mechanism that generates the chirp, which is explained in the supplement~\cite{supp}.}\label{fig.pm}
\end{figure}

As shown in Figure~\ref{fig.pca}, one can tell by inspecting the orientation of the fringes that the component of the rotation vector in the $xy$-plane points to 45~$^{\circ}$ with respect to the image axis. However, the fringes induced by clockwise (CW) or counter-clockwise (CCW) rotations appear parallel in their respetive images. The CW and CCW rotations are distinguished by scanning the acceleration phase and observing in which direction the fringes translate. The rotation fringes propagate like a plane wave traveling in the cold-atom cloud when the acceleration phase increases or decreases. The population ratio varies as $\cos{[\mathbf{k}_\Omega\cdot\mathbf{r}-(2\pi T^2)\alpha+\phi_0]}$, where $\alpha$ is the Raman laser chirp rate and $\phi_0$ is the phase due to Raman lasers and a homogeneous acceleration. Because the phase gradients generated by opposite rotations have opposite sign, the rotation fringes move in opposite directions when we scan the acceleration phase by scanning the chirp rate. 

When the rotation is large enough to generate one fringe across the cloud, the translational movement of the rotation fringes gives a measurement of the acceleration. Figure~\ref{fig.pm} shows a plot of the offset phase extracted from the 2D fits as a function of the chirp rate. In this offset phase vs. chirp rate plot, we fit the data with a line and keep the slope as a fixed parameter with the value of $-2\pi T^2$ and $T=8$~ms. The sensitivity of the acceleration measurement is calculated from the acceleration phase fluctuations, $\delta\Phi$, with $\delta g/g=\delta \Phi/(gk_{\mathrm{eff}}T^2)$ \cite{Mazzoni.2015, McGuinness.2012}. In our case, we interpret the root-mean-square (RMS) of the residuals of the linear fit as $\delta \Phi$. In Figure~\ref{fig.pm}, $\delta \Phi$ is 0.369~rad with 0.2~s between data points, corresponding to a fractional acceleration sensitivity $\delta g/g$ of $1.6\times10^{-5}\surd\mathrm{Hz}$ when $T=8$~ms. The sensitivity is currently limited by the Raman laser phase noise and the vibration noise~\cite{supp}.

\section{Rotation vector sensitivity}\label{sec.allan}
\begin{figure*}
\includegraphics[width=6in]{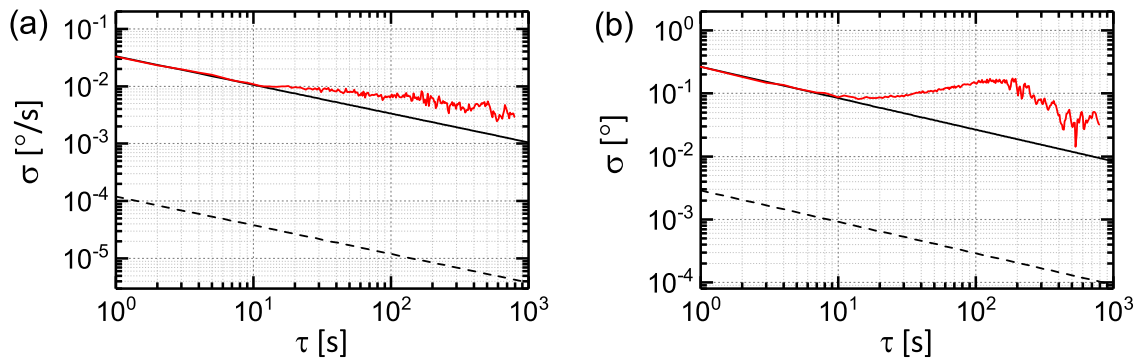}
\caption{(Color online.) The sensitivity of the (a) rate and (b) direction of the rotation vector measurement in the plane perpendicular to the direction of the Raman laser beams. The black solid lines are $\tau^{-1/2}$ fits from $\tau=1$ to 10~s. The black dashed lines are the performances of an ideal point source limited by atom shot noise, estimated with total atom number $4\times10^6$ coherently interacting with the Raman interrogation lasers, image size $1.8\times 1.8$~mm$^2$, $T=8$~ms, and no dead time. The simulated rotation is $3.04~^\circ/$s in the $+x$ direction. The average of measured rotation rate is $2.44~^\circ/$s.}\label{fig.sensitivity}
\end{figure*}

Figure~\ref{fig.sensitivity} shows the Allan deviation plots of the magnitude and the direction measurements of the rotation vector in the plane perpendicular to the direction of the Raman laser beams from a data set with a data rate of 1~Hz and duration of 4000~s. For both quantities $\sigma$ goes down as $\tau^{-1/2}$ from $\tau=1$ to 10 seconds. The intercept of the $\tau^{-1/2}$ power fit lines at $\tau=1$~s gives an estimate of the sensitivity of measuring a rotation vector in the plane perpendicular to the Raman laser beams: the sensitivity for the rate is $0.033~^{\circ}/\mathrm{s}$ and the sensitivity for the direction is $0.27~^{\circ}$ at 1~s of averaging time. 
 
In this data set, the experimental repetition rate is 10~Hz. In every 1~s, we record 10 rotation fringe images as the acceleration phase is scanned by 4~rad over the 10 images while the chirp rate is scanned from $-25.086$ to $-25.076$~kHz/ms. We process the group of ten images with PCA and reconstruct these images with the sine-like and cosine-like PCs. All images are cropped to $1.8\times1.8$ mm$^2$ ($70\times70$ pixels$^2$). By 2D fitting each reconstructed image we obtain ten rotation rates and ten rotation directions. The average of the ten rotation rates (rotation directions) are used; as a result, we have a list of 4000 rotation rates (rotation directions) with a data rate of 1~Hz as the input for the Allan deviation calculation. 

The rotation measurements are based on the phase gradient in the cold-atom cloud, and since the laser phase noise and vibrations (parallel to ${\mathbf{k}}_\mathrm{eff}$) create common-mode noise across the cloud that affects only the offset of the phase, the rotation sensitivity should be independent of those contributions to first order. The contributions to the noise in our rotation measurement are still under study. We estimate the contributions from the rotating mirror and the uncorrelated vibration of our floating optical table to be $2\times 10^{-3}$~$^{\circ}/\mathrm{s}$ and $5\times 10^{-5}$~$^{\circ}/\mathrm{s}$ at 1~s of averaging time, respectively. The hump in the direction plot is likely due to a slow oscillation in the raw data. This could be produced, for example, through the voltage control of the piezo-actuated platform that simulates the rotation and/or thermal effects.

The average of the measured rotation rate is 92~\% of the simulated rotation rate in Figure 3, 79~\% in Figure 4, and 80~\% in Figure 5, because of the finite size of the initial cloud~\cite{Hoth.2016}. The final cloud is 2.2 times bigger than the initial cloud in these measurement and we expect to see $1-\sigma_0^2/\sigma_f^2=0.79$ of the simulated rotation assuming a Gaussian cloud shape. The discrepancy in the measurement could come from the Gaussian shape approximation, fluctuations in cloud shape and size, or correlations between the position and velocity in the initial cloud when it is released from the MOT or molasses. Additional errors may arise from the calibration of the imaging system and the calibration of the simulated rotation. 

\section{conclusion and outlook}\label{sec.conclusion}

Based on the technique of PSI, we have demonstrated multi-axis inertial sensing: acceleration in the direction of the Raman laser beams and the component of the rotation vector in the plane perpendicular to that direction. The sensitivity of our present experiment is primarily constrained by the short Raman interrogation time ($T=8$~ms), technical noise, the initial size of the cold-atom cloud, and the measurement dead time. 

The finite initial size of the cold-atom cloud causes systematic errors in the rotation vector measurement. While Bose--Einstein condensates (BECs) provide more point-like cold-atom sources, present setups involving BECs are challenging to implement with low SWaP and high bandwidth \cite{Rudolph.2015}. Nevertheless, stabilizing the cloud shape may allow us to control and to minimize such systematic errors. With a Gaussian cloud shape, the cloud width in the direction of the Raman laser beams does not affect the rotation vector measurement on the plane perpendicular to that direction; only the cloud shape in the plane does. We could use a one-dimensional optical lattice trap in the direction of the Raman laser beams to initialize the cloud shape in the transverse direction. The density profile of the cloud released from a lattice trap relies on the profile and intensity of lattice laser beam, which can be adjusted and feedback controlled. 

The PSI instrument may be used very naturally as a gyrocompass. In our experiment, the direction of the Raman laser beams is parallel to $\mathbf{g}$ and the cold-atom cloud is imaged in the plane normal to $\mathbf{g}$; therefore the atoms sense the projection of the rotation vector into the plane tangent to the surface of the Earth at the sensor location. The component of Earth's rotation in this plane points to the geographic north. Hence, the direction of the rotation fringes due to Earth's rotation will point to the geographic north and the number of the fringes will be proportional to the cosine of the latitude of the sensor location.

The PSI gyroscope is analogous to a mechanical gyroscope made of a spinning rotor. When the rotor spins about the $z$-axis, it senses the component of torque in the $xy$-plane. In PSI, when the direction of the Raman laser beams is along the $z$-axis, the atoms sense the rotation component in the $xy$-plane. The two-dimentional sensitivity of PSI may have applications in detecting time-varying rotation vectors, which is needed to measure a precession. In a simple case where the rotation vector traces a cone centered about the $z$-axis, the precession can be measured by tracking the rotation component in the $xy$-plane. Tracking the direction of a rotation vector is crucial for measuring relativistic precessions \cite{Jentsch.2004}, as demonstrated by the space experiment Gravity Probe B~\cite{GPB}, in which state-of-the-art gyroscopes made of spinning rotors were deployed. 

\section*{Acknowledgements}
We thank Mark~A.~Kasevich for helpful discussions. We thank Rodolphe~Boudot, James~P.~McGilligan, and Moshe~Shuker for their comments on the manuscript. A.~H. was supported for this work under an NRC Research Associateship award at NIST. This work was funded by NIST, a U.S. government agency, and it is not subject to copyright.

\nocite{apsrev41Control}
\bibliographystyle{apsrev4-1}
\bibliography{bibliography}

\end{document}


\title{Supplemental material of multi-axis atom interferometer gyroscope with a single source of atoms}
\date{\today}

\author{Yun-Jhih~Chen\textsuperscript{1,2}}
\author{Azure~Hansen\textsuperscript{1}}
\author{Gregory~W.~Hoth\textsuperscript{1,2}}
\altaffiliation{Present address: Department of Physics, University of Strathclyde, Glasgow, United Kingdom}
\author{Eugene~Ivanov\textsuperscript{1,3}}
\author{Bruno~Pelle\textsuperscript{1,2}}
\altaffiliation{Present address: MUQUANS, Institut d'Optique d'Aquitaine, rue Fran\c{c}ois Mitterrand, 33400, Talence, France. }
\author{John~Kitching\textsuperscript{1}}
\author{Elizabeth~A.~Donley\textsuperscript{1}}

\affiliation{\textsuperscript{1}National Institute of Standards and Technology, Boulder, CO 80305}
\affiliation{\textsuperscript{2}University of Colorado Boulder, Boulder, CO 80309}
\affiliation{\textsuperscript{3}University of Western Australia, Perth, WA, Australia}

\maketitle

\section{laser system}
\begin{figure*}
\includegraphics[width=\textwidth]{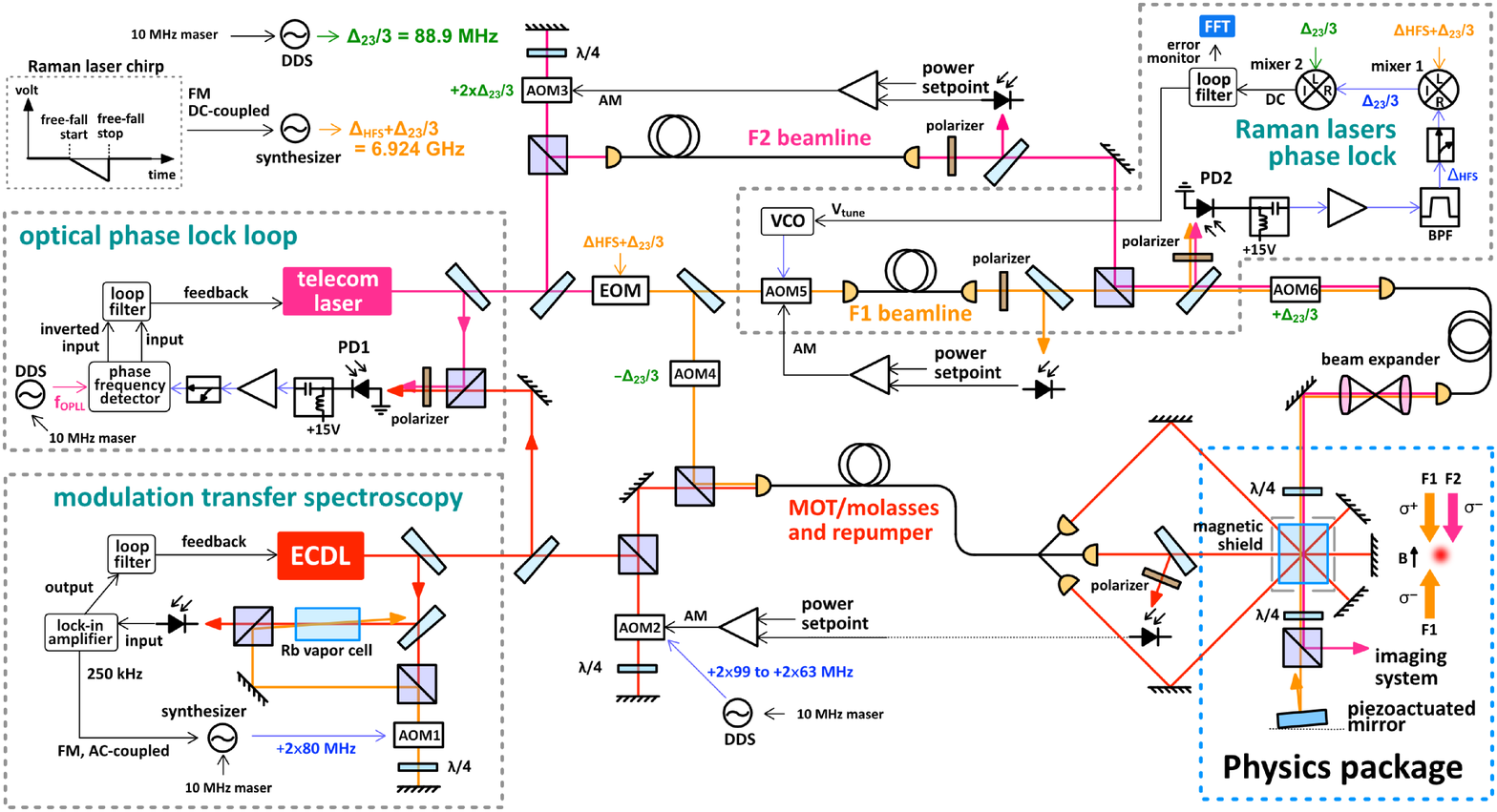}
\caption{Laser system. The ECDL laser is locked to a \textsuperscript{87}Rb saturated absorption line. The telecom laser is locked to the ECDL via an OPLL. The set point f\textsubscript{OPLL} cycles through four frequencies (as listed in Figure~\ref{figs.levels}). With the use of the EOM and AOMs, the system provides all laser frequencies required in our experiment. The scanning of the Raman laser chirp rate is done by varying the slope of the triangular waveform applied to the DC-coupled FM port of the synthesizer. AM: amplitude modulation. FM: frequency modulation.}\label{figs.schematic}
\end{figure*}

\begin{figure}
\includegraphics[width=3.2in]{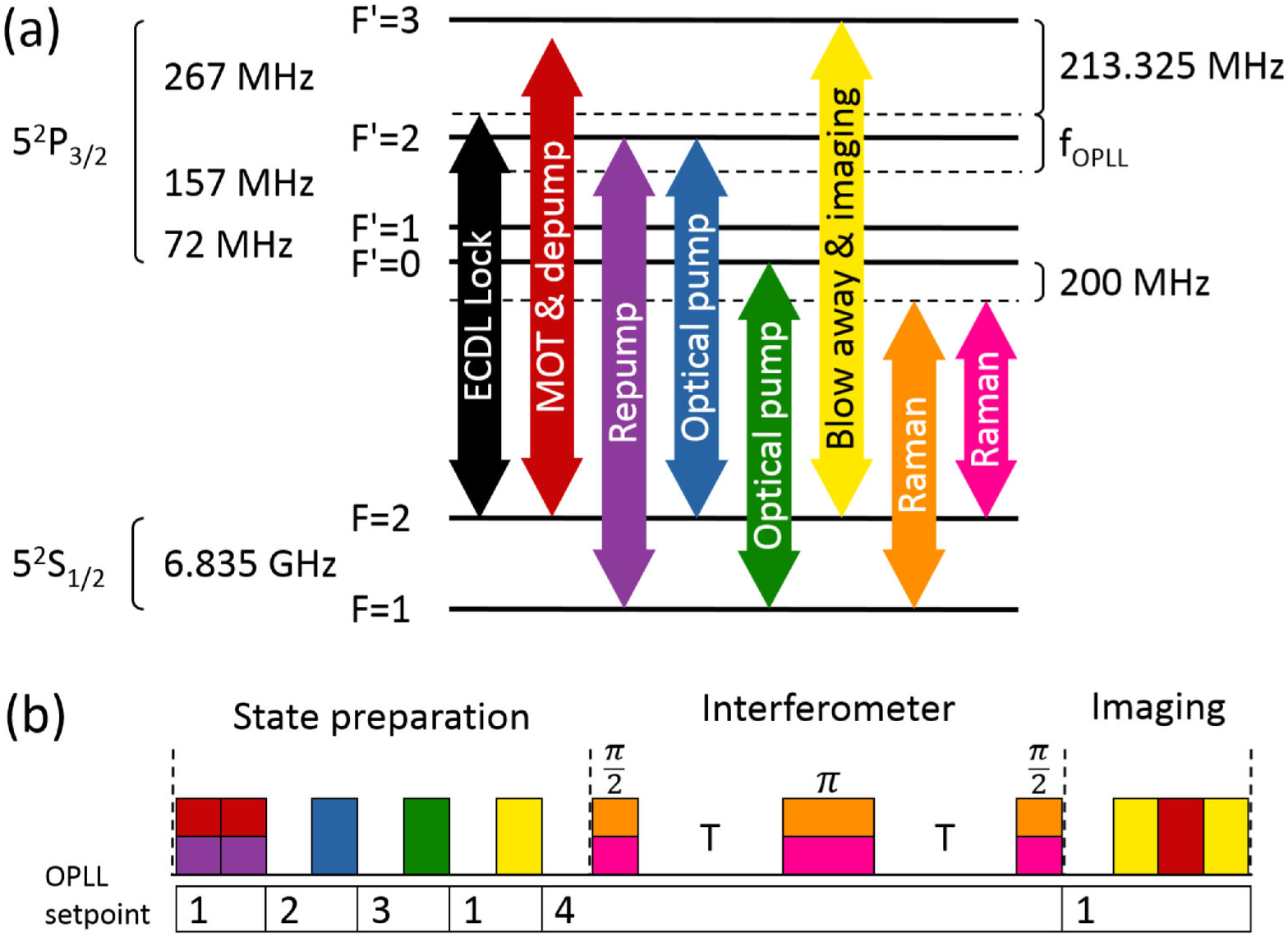}
\caption{Laser transition frequencies (a) and timing sequence (b). The black arrow shows the ECDL laser frequency when it is locked. The OPLL set point (f\textsubscript{OPLL}) cycles through four frequencies: (1) 53.325 (2) 319.975 (3) 549.140 (4) 749.140~MHz. The experimental repetition rate is 5~Hz or 10~Hz. The energy levels and timing sequence are not to scale.}\label{figs.levels}\end{figure}

Figure~\ref{figs.schematic} is a schematic of our laser system which provides the laser frequency components shown in Figure~\ref{figs.levels}. We use two commercially available 780~nm lasers. One is an external-cavity diode laser (ECDL) and the other is a frequency-doubled telecom laser which has a 20 kHz linewidth and output power up to 1~W. The ECDL, which is locked to a \textsuperscript{87}Rb saturated absorption line via modulation transfer spectroscopy \cite{msatspec}, provides the MOT cooling light. The telecom laser, locked to the ECDL via a heterodyne optical phase lock loop (OPLL) \cite{Hockel.2008}, provides the laser frequency components for all other frequencies in the experiment, including the repumper, state preparation lasers, Raman lasers, and the probe laser. The OPLL allows us to switch the telecom laser frequency over a range of several hundreds of MHz in a few hundreds of $\mu$s.

For modulation transfer spectroscopy, a double-pass acousto-optic modulator (AOM1 in Figure~\ref{figs.schematic}) shifts the pump frequency by +160~MHz, so the Doppler-free spectral lines move by $-80$~MHz. The ECDL is locked to the red-shifted crossover transition of  $F=2\to F'=2,3$. The lock signal is generated by modulating the frequency of the 80 MHz signal applied to AOM1 with a 250 kHz sine wave from the lock-in amplifier. The resulting frequency deviation is close to the half-linewidth of the saturated absorption resonance of $^{87}$Rb. Such a choice maximizes the frequency-to-voltage conversion efficiency of the modulation transfer spectrometer. Most of the ECDL power goes to another double-pass AOM in a cat's eye configuration \cite{CatEye} (AOM2 in Figure~\ref{figs.schematic}) that shifts the laser frequency close to the cycling transition for use in the MOT. The light is red-detuned from the cycling transition by $2.5~\Gamma$ during the MOT phase and is linearly chirped to $14.4~\Gamma$ during the molasses phase. The power of the cooling laser is actively stabilized and is linearly ramped down by a factor of ten during the molasses phase.

A small portion of the ECDL and the telecom laser light are combined onto a fast photodiode (PD1 in Figure~\ref{figs.schematic}) that measures the beat note between the two lasers. A digital phase frequency detector (DPFD) compares the beat frequency with a stable RF reference (f\textsubscript{OPLL}) and generates the OPLL lock signal. The telecom laser is phase-locked to the ECDL at a frequency that is below the lock point of the ECDL laser by f\textsubscript{OPLL} (Figure~\ref{figs.levels}). The range of f\textsubscript{OPLL} is limited by the maximal operating frequency of the DPFD (close to 1 GHz). The fractional power of the carrier in the beat note is greater than 99.73~\%, which corresponds to $2.7\times10^{-3}$ rad$^2$ in the residual phase fluctuation between the two lasers \cite{Hockel.2008}. 

Most of the telecom laser output is split into two beamlines, which we refer to as the ``F1'' and ``F2'' beamlines in Figure~\ref{figs.schematic}. The F1 (F2) beamline is used for transitions starting from the $F=1$ ($F=2$) manifold. The F1 beamline contains an electro-optic phase modulator (EOM) that creates sidebands at $\Delta_\text{HFS}+\Delta_{23}/3 = 6.924$ GHz, where $\Delta_\text{HFS}=6.835$ is the ground state hyperfine splitting frequency of \textsuperscript{87}Rb and $\Delta_{23}=266.650$~MHz is the difference between the $F'=2,3$ levels in the $5P$ states. The carrier is nulled by tuning the RF power applied to the EOM. The F2 beam and the +1 sideband form the Raman beams. The combinations among the other sidebands and the F2 beam are off-resonant by multiples of $\Delta_{23}/3$. 

Despite the fact that both Raman beams are extracted from the same laser, their phase difference, i.e., the Raman laser phase, fluctuates because of the physically different F1 and F2 beam paths. The voltage controlled oscillator (VCO) frequency applied to AOM5 in the F1 beamline is feedback-controlled around 88.9~MHz to stabilize the phase difference of the two Raman beams. The DC lock signal is generated by mixing down the beat note of the Raman beams with the two references: a microwave synthesizer, which drives the EOM and mixer 1, and a DDS that provides the common RF source $\Delta_{23}/3$ to AOM3, 4, and 6, and mixer 2. 

\section{Raman laser phase noise}
\begin{figure}
\includegraphics[width=\columnwidth]{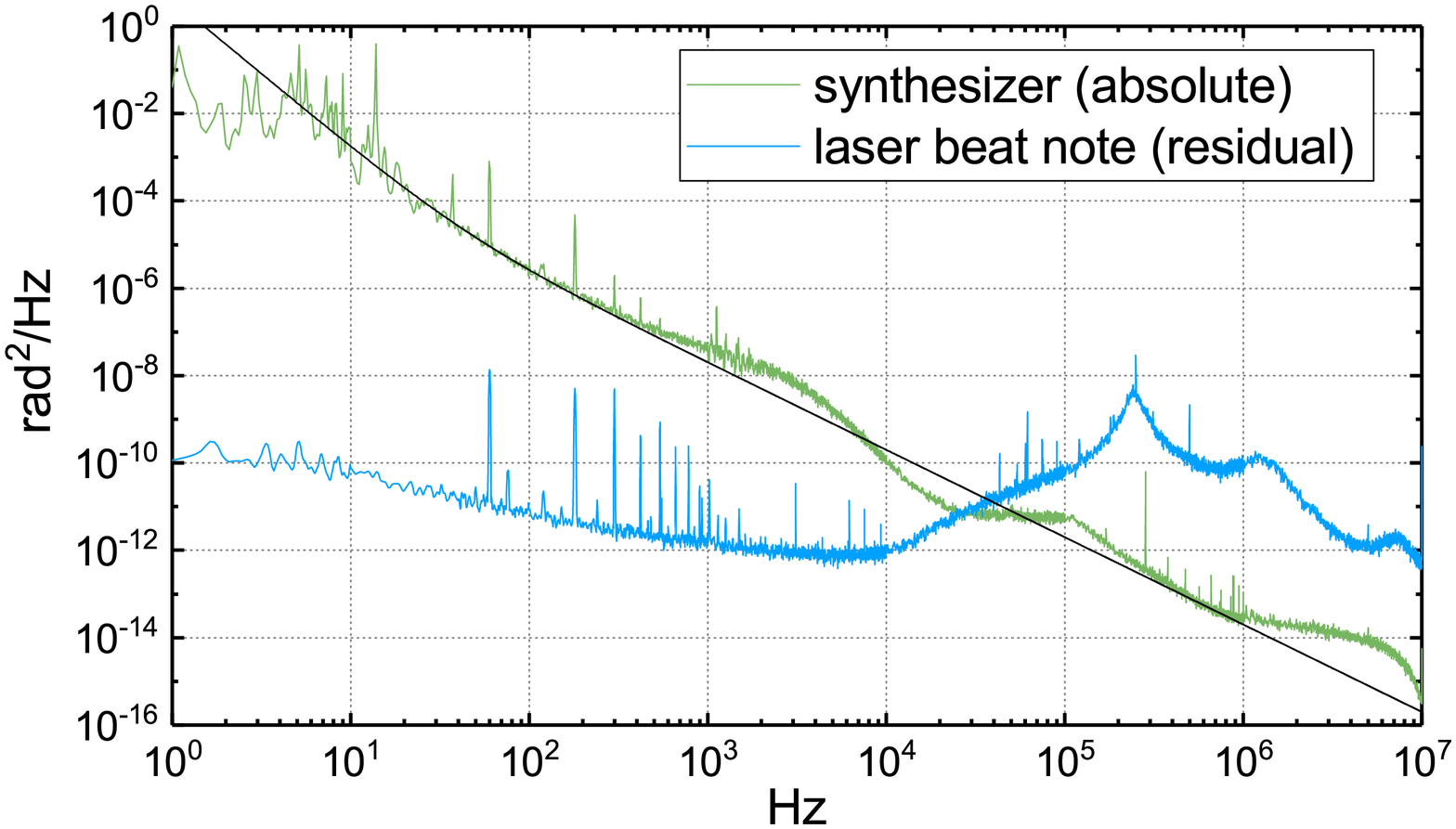}
\caption{Power spectral densities of the residual Raman laser phase noise and the absolute phase noise of the synthesizer used as a reference in the Raman laser phase lock. The spectra are taken when the DC-coupled FM port of the synthesizer is enabled. The black line is the power fit $4/f^{3.4}+0.02/f^2$ used in calculating the contribution from the synthesizer to the interferometer phase noise.}\label{figs.fft}
\end{figure}

Figure~\ref{figs.fft} shows the power spectral densities of the major phase noise sources present in our Raman lasers, including the residual Raman laser phase noise and the absolute phase noise of the microwave synthesizer used as one of the two RF references in phase locking the Raman laser beams. The residual Raman laser phase noise, measured from the beat frequency of the Raman laser beams when the phase lock is enabled, shows only the Raman laser phase fluctuation relative to the RF references. The Raman phase noise that affects the atoms is the residual Raman laser phase noise plus the phase noise of the the RF references. However, in the low frequency range 1--$10^4$~Hz that affects the interferometer phase the most, the contribution from the synthesizer dominates. Following the approach in reference \cite{Cheinet.2008}, with $T=8$~ms, $t_\pi=4.4~\mu$s, and an experimental repetition rate of 10~Hz, we estimate the instability of the interferometer phase arising from the Raman laser phase noise (due to the microwave synthesizer) to be 90~mrad at 1~s. The corresponding instability of $\delta g/g$ at 1~s is $8.91\times10^{-6}$.  In comparison, the phase instability due to the vibration (in the direction of the Raman lasers) of the mirror which retro-reflects only the Raman F1 laser is approximately 62~mrad at 1~s. The corresponding instability of $\delta g/g$ at 1~s is $6.14\times10^{-6}$.  The experiment is mounted on a floating optical table with no active vibration isolation.

\nocite{apsrev41Control}
\bibliographystyle{apsrev4-1}
\bibliography{bibliography}